\RequirePackage{lineno}
\documentclass[aps,prl,twocolumn,groupedaddress,amsmath,amssymb]{revtex4-1}

\usepackage{amssymb}
\usepackage{graphicx}
\usepackage{subfigure}
\usepackage{dcolumn}
\usepackage{bm}
\usepackage{epsfig}
\usepackage{xcolor}
\usepackage{enumerate}
\usepackage{booktabs}
\usepackage{hyperref}
\hypersetup{hidelinks,
	colorlinks=true,
	allcolors=blue,
	pdfstartview=Fit,
	breaklinks=true}
\usepackage{float}
\usepackage{multirow}

\begin{document}


\newcommand{\jpsi}{J/\psi}
\newcommand{\pip}{\pi^+}
\newcommand{\pin}{\pi^-}
\newcommand{\pio}{\pi^0}
\newcommand{\g}{\gamma}
\newcommand{\gev}{GeV/c$^2$}
\newcommand{\mev}{MeV/c$^2$}
\newcommand{\ar}{\rightarrow}
\newcommand{\ks}{K_S^{0}}
\newcommand{\etap}{\eta^\prime}

\newcommand{\BESIIIorcid}[1]{\href{https://orcid.org/#1}{\hspace*{0.1em}\raisebox{-0.45ex}{\includegraphics[width=1em]{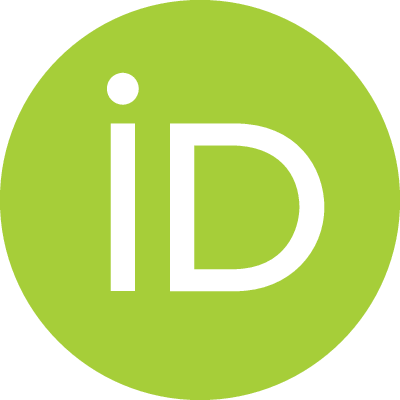}}}}

\title{\Large \boldmath \bf Observation of a strange $J^P=1^{-}$ resonance $K^{*}_1(1870)$ in $\chi_{c1}\to K^+K^-\eta^{\prime}$ decays}

\author{
  \begin{center}
M.~Ablikim$^{1}$\BESIIIorcid{0000-0002-3935-619X},
M.~N.~Achasov$^{4,c}$\BESIIIorcid{0000-0002-9400-8622},
P.~Adlarson$^{85}$\BESIIIorcid{0000-0001-6280-3851},
X.~C.~Ai$^{91}$\BESIIIorcid{0000-0003-3856-2415},
C.~S.~Akondi$^{32A,32B}$\BESIIIorcid{0000-0001-6303-5217},
R.~Aliberti$^{40}$\BESIIIorcid{0000-0003-3500-4012},
A.~Amoroso$^{84A,84C}$\BESIIIorcid{0000-0002-3095-8610},
Q.~An$^{80,67,\dagger}$,
M.~S.~Anderson$^{40}$\BESIIIorcid{0009-0008-1550-2632},
Y.~Bai$^{65}$\BESIIIorcid{0000-0001-6593-5665}
% ...
\\
\vspace{0.2cm}
(BESIII Collaboration)\\
\vspace{0.2cm} {\it
$^{1}$ Institute of High Energy Physics, Beijing 100049, People's Republic of China\\
% ...
$^{4}$ Budker Institute of Nuclear Physics SB RAS (BINP), Novosibirsk 630090, Russia\\
% ...
$^{32}$ INFN Sezione di Ferrara, (A)INFN Sezione di Ferrara, I-44122, Ferrara, Italy; (B)University of Ferrara, I-44122, Ferrara, Italy\\
% ...
$^{40}$ Johannes Gutenberg University of Mainz, Johann-Joachim-Becher-Weg 45, D-55099 Mainz, Germany\\
% ...
$^{65}$ Southeast University, Nanjing 211100, People's Republic of China\\
% ...
$^{67}$ State Key Laboratory of Particle Detection and Electronics, Beijing 100049, Hefei 230026, People's Republic of China\\
% ...
$^{80}$ University of Science and Technology of China, Hefei 230026, People's Republic of China\\
% ...
$^{84}$ University of Turin and INFN, (A)University of Turin, I-10125, Turin, Italy; (B)University of Eastern Piedmont, I-15121, Alessandria, Italy; (C)INFN, I-10125, Turin, Italy\\
$^{85}$ Uppsala University, Box 516, SE-75120 Uppsala, Sweden\\
% ...
$^{91}$ Zhengzhou University, Zhengzhou 450001, People's Republic of China\\

\vspace{0.2cm}
$^{\dagger}$ Deceased\\
% ...
$^{c}$ Also at the Novosibirsk State University, Novosibirsk, 630090, Russia\\
% ...

}

  \end{center}
}

\date{\today}

\begin{abstract}

A partial wave analysis of the process $\psi(2S)\to \gamma\chi_{c1}$, $\chi_{c1}\to K^{+}K^{-}\eta^{\prime}$, with the $\eta^{\prime}$ meson reconstructed in the $\pi^+\pi^-\gamma$ and $\pi^+\pi^-\eta$ decay modes, is performed using a sample of $(2712.4 \pm 14.3)\times10^{6}$ $\psi(2S)$ events collected with the BESIII detector.  In the $K\eta^{\prime}$ system, a resonant contribution with quantum numbers $J^{P}=1^{-}$ is observed for the first time and is denoted as the $K_1^*(1870)$ meson. Its statistical significance is $11.0\sigma$, and its mass and width are measured to be $M=(1869.0 \pm 14.6\pm 18.6$) MeV/$c^{2}$ and $\Gamma=(167.3 \pm 34.0\pm 24.2$) MeV, respectively. The branching fraction of the $\chi_{c1}\to K^{\mp}K_1^*(1870)^{\pm}\to K^{+}K^{-}\eta^{\prime}$ decay chain is determined to be $(1.48 \pm 0.30\pm 0.33)\times10^{-5}$. The branching fraction of the $\chi_{c1}\to K^{+}K^{-}\eta^{\prime}$ decay is measured to be $(7.90 \pm 0.48)\times10^{-4}$. Several intermediate scalar, vector, and tensor states are required to provide a good description of the data. These results provide new input for understanding strange-meson spectroscopy and investigating possible hybrid-state candidates.
\end{abstract}

\maketitle

The search for hadrons with exotic quantum numbers is one of the central topics in hadron spectroscopy, because such states offer a direct opportunity to probe  the non-perturbative dynamics of Quantum Chromodynamics (QCD). In the conventional quark model, mesons are described as quark--antiquark ($q\bar q$) bound states, and their allowed $J^{PC}$ quantum numbers are therefore strongly constrained. States with $J^{PC}=1^{-+}$ are particularly interesting because they cannot be formed by a simple $q\bar{q}$ configuration, for which $P=(-1)^{L+1}$ and $C=(-1)^{L+S}$, and therefore constitute a clear signature of exotic matter~\cite{Meyer:2015gp}. Among the most compelling interpretations of such exotic states are hybrid mesons, in which an excited gluonic field contributes explicitly to the hadron structure~\cite{Meyer:2015gp,KLEMPT20071}. Establishing the existence of $1^{-+}$ hybrids is thus of great importance for understanding gluonic excitations, color confinement, and the full hadron spectrum predicted by QCD.

Lattice QCD predicts that the lightest hybrid meson with exotic quantum numbers $1^{-+}$ should lie in the mass range of  1.7 to 2.1 GeV/$c^{2}$~\cite{lightest_hybrid, hybrid}. Experimental searches for such states have yielded several important candidates. The CLEO Collaboration reported the first evidence for the $\pi_{1}(1600)$ meson in the $\pi\eta^\prime$ system in the decay $\chi_{c1}\to \pi^+\pi^-\eta^\prime$~\cite{pi1600}, which has recently been confirmed by the BESIII experiment using the same final state~\cite{BESIII:2026ebb}. In addition, the BESIII Collaboration observed an isoscalar exotic candidate, $\eta_{1}(1855)$, in the process $J/\psi\to \gamma\eta\eta^\prime$~\cite{eta1855}. Earlier evidence for other $1^{-+}$ candidates, including the $\pi_{1}(1400)$ and $\pi_{1}(2015)$, was also reported by the E852 Collaboration~\cite{pi1400,pi2015}.

These experimental discoveries have significantly advanced experimental efforts to establish the light $1^{-+}$ hybrid nonet predicted by QCD~\cite{nonet_1, nonet_2}.  
In particular, the observations of the $\pi_{1}(1600)$~\cite{pi1600,BESIII:2026ebb} and the $\eta_{1}(1855)$~\cite{eta1855} mesons provide important candidates for the isovector and isoscalar members of this nonet.  However, a crucial missing piece is still the strange partner. Its observation is essential not only for completing the nonet structure, but also for discriminating among different theoretical interpretations of the exotic spectrum. In particular, the scenario that identifies the $\eta_{1}(1855)$ meson as the lower-mass isoscalar state predicts the existence of a strange hybrid with a mass around 1.83--1.90~GeV/$c^{2}$~\cite{nonet_1}. Such a state is expected to couple to the $K\eta^\prime$ system.

In this context, the decay $\chi_{c1}\to K^+ K^- \eta^\prime$ provides a particularly promising laboratory to search for a strange hybrid candidate. The decay also offers an excellent opportunity to investigate scalar and tensor mesons in the $K^+K^-$ system, as well as conventional strange mesons, as demonstrated in a previous BESIII study~\cite{previouswork}. The decay is studied via a partial wave analysis (PWA) of the decay where the initial state is produced via the $\psi(2S)\to\gamma\chi_{c1}$ decay. With a data sample of $(2712.4\pm14.3)\times10^6$ $\psi(2S)$ events~\cite{ref::psip_event}, which is about five times larger than the subsample used in Ref.~\cite{previouswork}, together with excellent performance in photon detection, kaon identification, and $\eta^\prime$ reconstruction, the BESIII experiment is uniquely well suited to search for the possible exotic states with strangeness in the $K\eta^\prime$ system and to test the flavor-multiplet structure of light hybrid mesons. The BESIII detector is described in detail in Ref.~\cite{BESIIIdetector}. Monte Carlo (MC) simulated data samples are produced with a {\sc geant4}-based~\cite{geant4} software package, which includes the geometric description of the BESIII detector and the detector response. They are used to determine detection efficiencies and to estimate backgrounds.

In the analysis of the $\chi_{c1}\to K^+K^-\eta^{\prime}$ decay, the $\eta^{\prime}$ candidates are reconstructed through decays to $\pi^+\pi^-\gamma$ (Mode~I) and $\eta \pi^+\pi^-$ with $\eta\to\gamma\gamma$ (Mode~II) final states.  Since the $\chi_{c1}$ is selected from the 
$\psi(2S)$ radiative decays, the corresponding final states are 
$\gamma\gamma K^+K^-\pi^+\pi^-$ and  $\gamma\gamma\gamma K^+K^-\pi^+\pi^-$, respectively, for the two modes. The selection criteria for the charged tracks, photon showers, and particle identification (PID) for the kaons and pions follow those used in the previous BESIII analysis of $\chi_{cJ}\to K^+K^-\eta^{\prime}$~\cite{previouswork}.

Candidate events are required to contain four good charged tracks identified as $K^+K^-\pi^+\pi^-$ and at least two photons for Mode~I or three photons for Mode~II. To improve the momentum and energy resolution and to suppress background contributions, a four-constraint (4C, energy-momentum constraint) kinematic fit is applied to Mode~I, while a five-constraint (5C) kinematic fit is used for Mode~II with the invariant mass of the photon pair assigned to the $\eta$ candidate constrained to the known $\eta$ mass~\cite{pdg}. 
If more than two (three) photon candidates are present in Mode~I (Mode~II), all possible combinations are considered, and the one with the smallest $\chi^2_{\rm4C}$ ($\chi^2_{\rm5C}$) is retained. The selected combination is required to satisfy $\chi^2_{\rm4C}<40$ for Mode~I and $\chi^2_{\rm5C}<100$ for Mode~II. To further suppress background events with either one more or one less photon for Mode~I, the $\chi^2_{\rm4C}$ value under the signal hypothesis $\psi(2S)\to \gamma K^+K^-\pi^+\pi^-\gamma$ is required to be smaller than those obtained under the corresponding background hypotheses.

The  $\eta^\prime$ candidates are selected by requiring
$|M(\pi^+\pi^-\gamma)-m(\eta^\prime)|<15$~MeV/$c^{2}$ for Mode~I and $|M(\pi^+\pi^-\eta)-m(\eta^\prime)|<10$~MeV/$c^{2}$ for Mode~II, where $m(\eta^\prime)$ is the known mass of $\eta^\prime$~\cite{pdg}.  
To suppress backgrounds from $J/\psi$ decays, events with a reconstructed $K^+K^-\pi^+\pi^-$ or $K^+K^-\gamma\gamma$ invariant mass falling within the $J/\psi$ mass window ($|M(K^+K^-\pi^+\pi^-)-m(J/\psi)|<22$ MeV/$c^{2}$ or $|M(K^+K^-\gamma\gamma)-m(J/\psi)|<8$ MeV/$c^{2}$) are rejected, where $m(J/\psi)$ is the known mass of $J/\psi$~\cite{pdg}.  
In addition, to suppress backgrounds containing $\pi^0$ meson decays, events with a $\gamma\gamma$ invariant mass within the $\pi^0$ mass window ($|M(\gamma\gamma)-m(\pi^0)|<15$ MeV/$c^{2}$) are  vetoed, where $m(\pi^0)$ is the known mass of $\pi^0$~\cite{pdg}.
After vetoing possible $J/\psi$ and $\pi^0$ contributions, the remaining background sources are found, according to studies based on inclusive MC simulations, to arise mainly from $\psi(2S)\to\gamma\chi_{cJ}$ with $\chi_{cJ}\to (\pi^0)K^+K^-\pi^+\pi^-$ in Mode~I, and from non-$\eta^\prime$ processes such as $\chi_{cJ}\to K^+K^-\pi^+\pi^-\eta$ in Mode~II. These backgrounds can be effectively described using the $\eta^\prime$ sideband regions.

\begin{figure}[htbp]
    \centering
    \subfigure{
    \includegraphics[width=3.in]{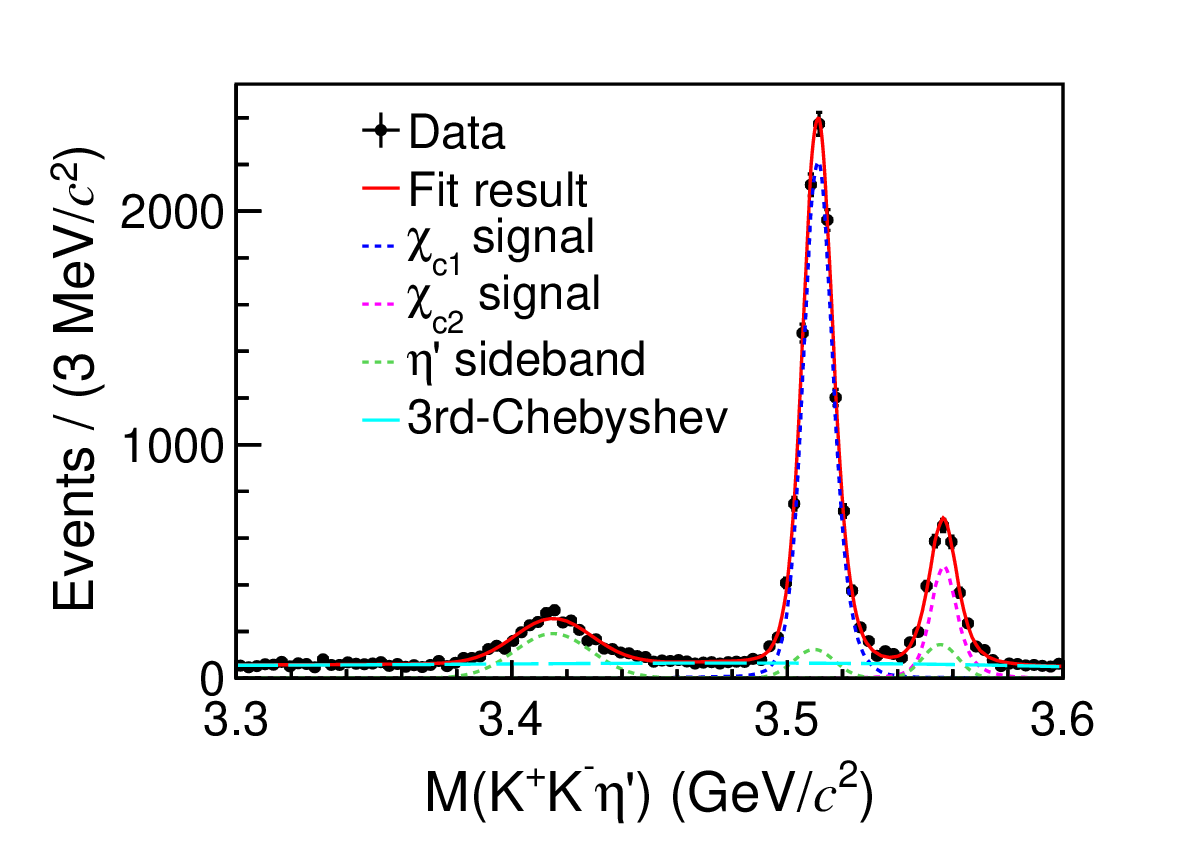}
    \put(-45,125){ \large~(a)}
    \label{pipigamma}
    }
    \subfigure{
    \includegraphics[width=3.in]{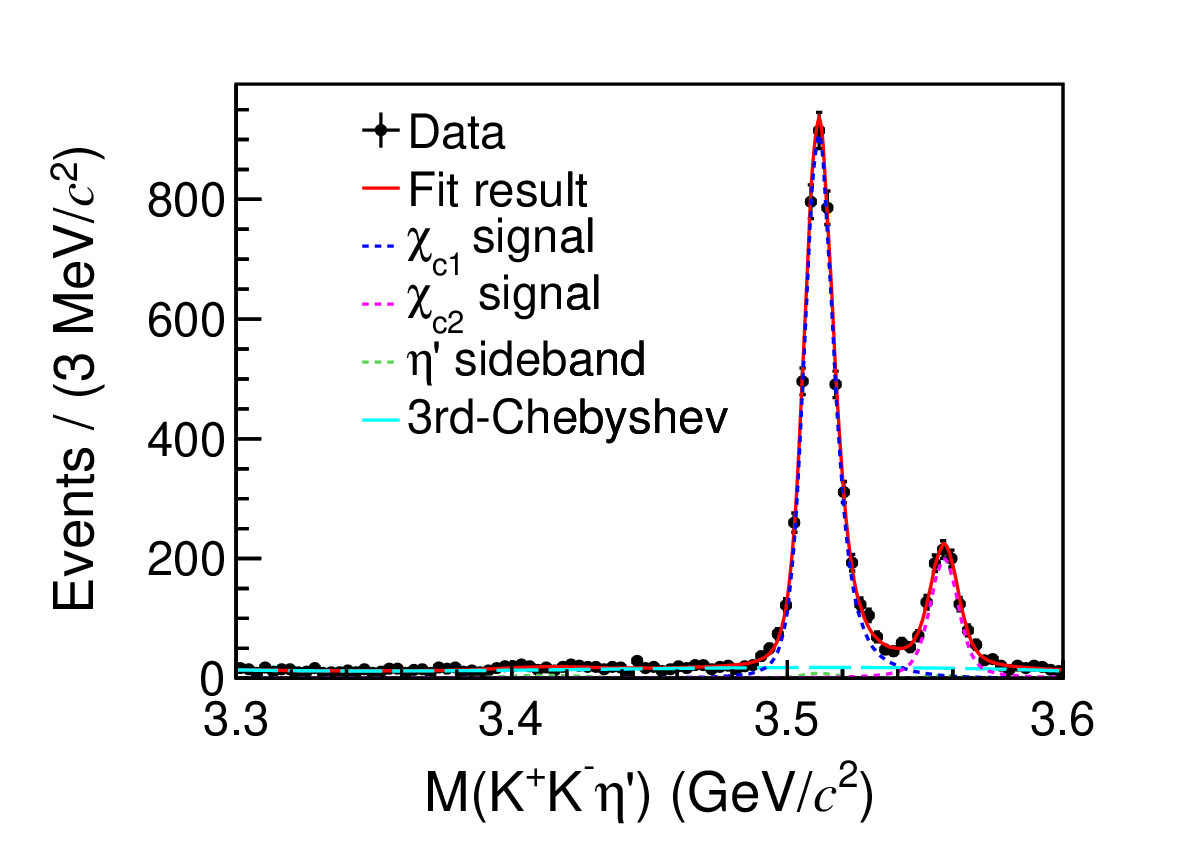}
    \put(-45,125){ \large~(b)}
    \label{pipieta}
    }
    \caption{Invariant mass spectra of  $K^+K^-\eta^{\prime}$ for (a) $\eta^{\prime}\to\pi^+\pi^-\gamma$ and (b)  $\eta^{\prime}\to\pi^+\pi^-\eta$.}
    \label{fit_kketap}
\end{figure}

After applying the above requirements, the $K^+K^-\eta^\prime$ invariant mass spectra for Mode~I and Mode~II are shown in Figs.~\ref{pipigamma} and \ref{pipieta}, respectively. Clear $\chi_{c1}$ and $\chi_{c2}$ signals are observed in both modes. 
In Mode~I, an additional enhancement is visible in the $\chi_{c0}$ mass region, which is attributed to background events from $\chi_{c0}\to K^+K^-\pi^+\pi^-$.
To extract signal yields, unbinned maximum likelihood fits are performed to the $M(K^+K^-\eta^{\prime})$ distribution in the $\chi_{cJ}$ signal region for both $\eta^{\prime}$ decay modes, as illustrated in Fig.~\ref{fit_kketap}.  
In the fits, the $\chi_{cJ}$ signals are modeled using the shapes from simulation convolved with Gaussian functions to account for possible resolution differences between data and simulation. The peaking backgrounds are described by the $\eta^{\prime}$ sideband events 
with their yields fixed according to fits to the $M(\pi^+\pi^-\gamma)$ or $M(\pi^+\pi^-\eta)$ distributions. The non-peaking backgrounds are parameterized by a third-order Chebyshev polynomial.

The signal yields for $\chi_{c1}\to K^+K^-\eta^\prime$ are $10786\pm116$ and $4654\pm75$ for Mode~I and Mode~II, respectively.
The branching fraction using both modes is determined to be
$(7.90 \pm 0.48)\times 10^{-4}$, taking into account the  number of $\psi(2S)$ events~\cite{ref::psip_event}, the relevant intermediate branching fractions from the PDG~\cite{pdg}, and the detection efficiencies $17.18\%$ for Mode~I and $13.44\%$ for Mode~II,  which are estimated from signal MC simulations based on the PWA results described below. 

Based on the open-source framework TF-PWA~\cite{TF-PWA}, a combined PWA is performed for the decays of $\psi(2S)\to\gamma\chi_{c1},\chi_{c1}\to K^+K^-\eta^{\prime}$ for the above two modes. 
A total of 11550 and 4617 events are selected within the $\chi_{c1}$ signal region, defined as $|M(K^+K^-\eta^\prime)-m(\chi_{c1})|<15$~MeV/$c^{2}$ and $<18$~MeV/$c^{2}$ for Mode~I and  Mode~II, respectively, where $m(\chi_{c1})$ is the $\chi_{c1}$ known mass~\cite{pdg}. 
The corresponding background contamination ratios are estimated to be 13.6\% and 5.9\% for Mode~I and Mode~II, respectively, which are evaluated using two-dimensional mass sideband regions in the $\eta^{\prime}$ and $\chi_{c1}$ invariant mass spectra and subtracted from the total likelihood value in the PWA fit. 

To construct the full decay amplitude, the helicity formalism combined with the isobar model is employed, in which the three-body decay is described as a sequence of quasi-two-body processes:
$\chi_{c1}\to X^{\pm}K^{\mp}$, followed by $X^{\pm}\to K^{\pm}\eta^\prime$, and 
$\chi_{c1}\to Y\eta^\prime$, followed by $Y\to K^+K^-$. Here, $X$ collectively denotes all intermediate resonant states decaying to the $K^{\pm}\eta^\prime$ final state, while $Y$ is reserved for intermediate states decaying to $K^+K^-$.
Most resonances are parameterized by relativistic Breit--Wigner propagators, unless otherwise specified. The $f_0(980)$ and $K_0^*(1430)$ resonances are described using the Flatt\'e formula~\cite{flatte}. The amplitudes for the charge-conjugate processes are constrained to share the same parameters.

To determine the final partial-wave composition for the decay $\chi_{c1}\to K^+K^-\eta^\prime$, a systematic resonance-selection procedure is carried out within the PWA framework. The main goal is to identify the minimal but sufficient set of intermediate two-body amplitudes that can describe the data with high statistical significance, while avoiding unnecessary model complexity.

In this case, all kinematically allowed resonances with quantum numbers $J^{P}=0^+$, $1^-$, $2^+$ for $X^{\pm}\to K^{\pm}\eta^\prime$ and $J^{P}=0^{++}, 2^{++}$ for $Y\to K^+K^-$ listed in the PDG~\cite{pdg} are considered as possible intermediate contributions. All relevant combinations of such amplitudes are tested in the fit. For each resonance, its statistical significance is evaluated from the change in the optimal negative log-likelihood value when the resonance is included in or removed from the fit, taking into account the corresponding change in the number of free parameters. Under the assumption that the likelihood-ratio test follows a $\chi^2$ distribution, the significance of each contribution is determined quantitatively. A PDG-based optimized solution is then obtained by retaining only those amplitudes whose significances exceed $5\sigma$.

After that, the possible additional contributions beyond the PDG-based solution are investigated. For this purpose, extra resonances with $J^{P(C)}=0^{+(+)}$, $1^-$, or $2^{+(+)}$ are introduced one by one into the fit, and the improvement in fit quality is examined. This study reveals that an additional $1^-$ structure in the $K\eta^\prime$ system around 1.8~GeV/$c^2$ is strongly favored by the data. By contrast, all other additional contributions lead to significances below $5\sigma$ and are therefore not retained. This observation motivates the inclusion of an extra $K_1^*$ resonance in the baseline solution.

The baseline amplitude set is then re-optimized after including this additional $K_1^*$ state, with its mass and width treated as free parameters in the fit. Since the introduction of a new resonance can affect the significance of other amplitudes through interference and correlations, all previously retained resonances are re-evaluated under the new fit hypothesis. States with updated significances below $5\sigma$ are removed. Different spin-parity assignments for the new state are also tested. The statistical significances for the $0^+$, $1^-$, and $2^+$ hypotheses are found to be $3.2\sigma$, $11.0\sigma$, and $3.7\sigma$, respectively, clearly favoring the $J^{P}=1^-$ assignment.

\begin{figure*}[htbp]
	\centering
    \subfigure{
    \includegraphics[width=2.in]{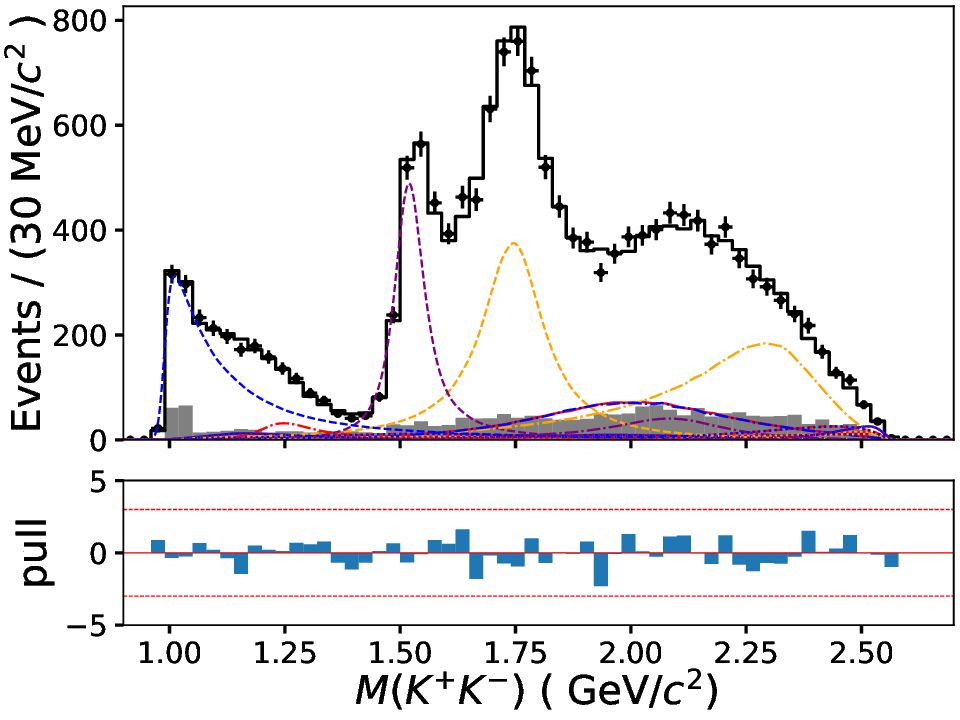}
    \put(-40,95){ \large~(a)}
    \label{fit_Mkpetap_mode1}
    }
    \hspace{-0.1in}
    \subfigure{
    \includegraphics[width=2.in]{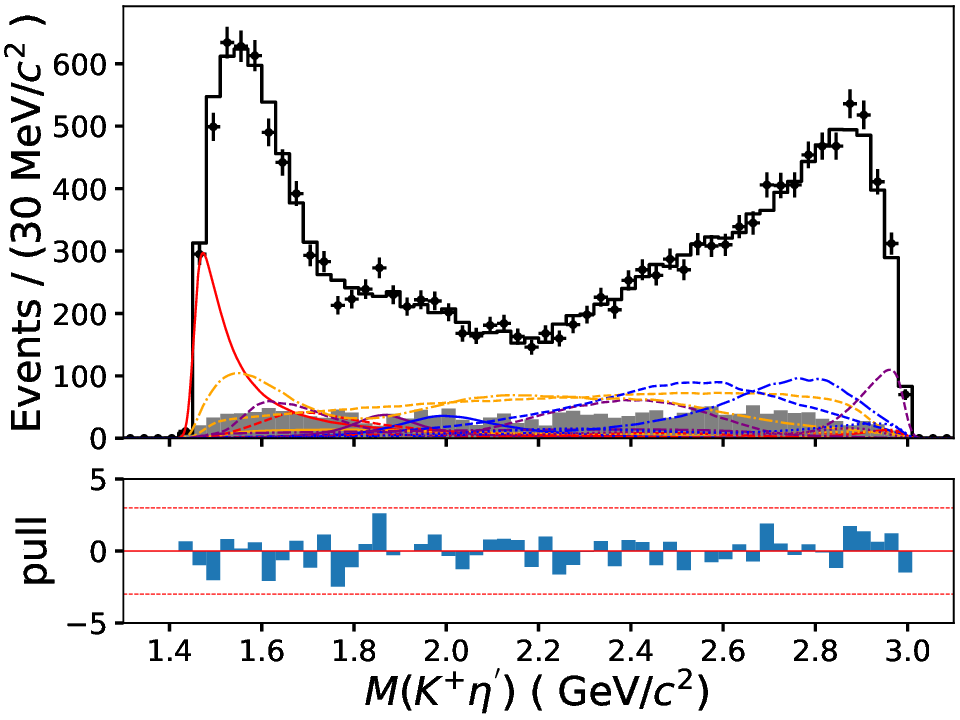}
    \put(-40,95){ \large~(b)}
    \label{fit_Mkmetap_mode1}
    }
    \hspace{-0.1in}
    \subfigure{
    \includegraphics[width=2.in]{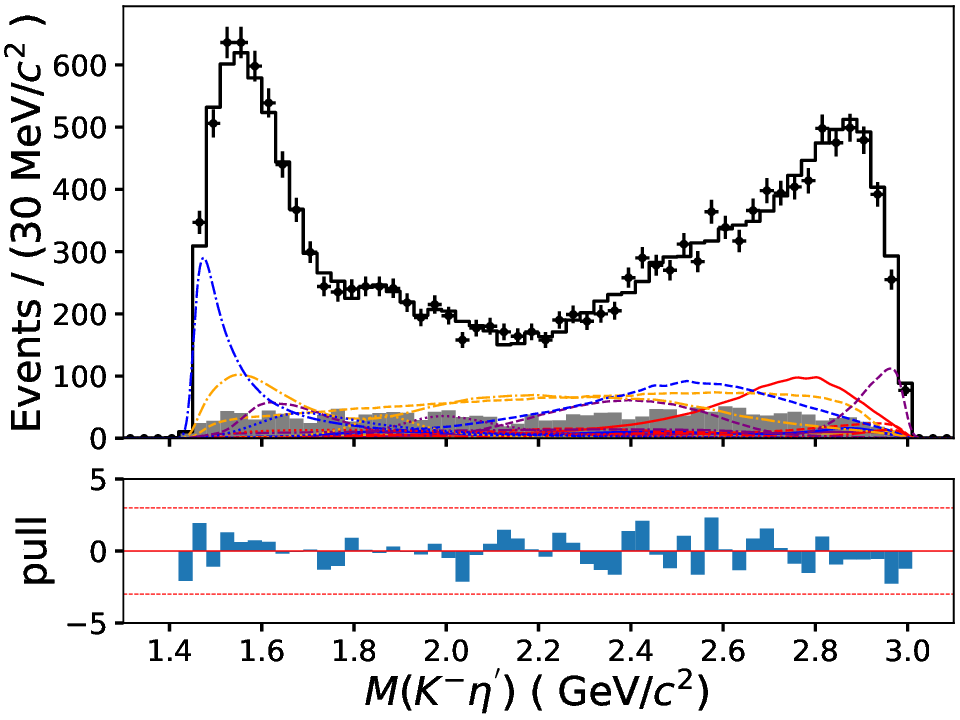}
    \put(-40,95){ \large~(c)}
    \label{fit_M2k_mode1}
    }
    \hspace{-0.1in}
\subfigure{
\raisebox{0.30in}{
\includegraphics[width=0.7in]{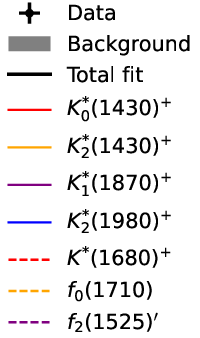}
}
}

    \subfigure{
    \includegraphics[width=2.in]{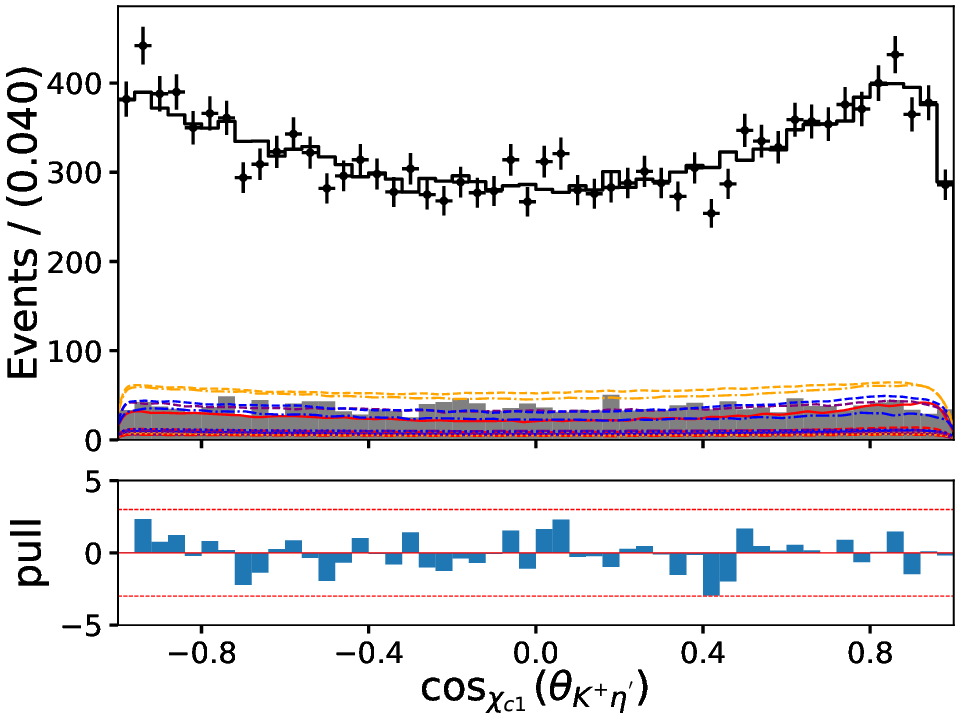}
    \put(-40,95){ \large~(d)}
    \label{fit_com_Rchic1_R_Kpetap_cos(beta)}
    }
    \hspace{-0.1in}
    \subfigure{
    \includegraphics[width=2.in]{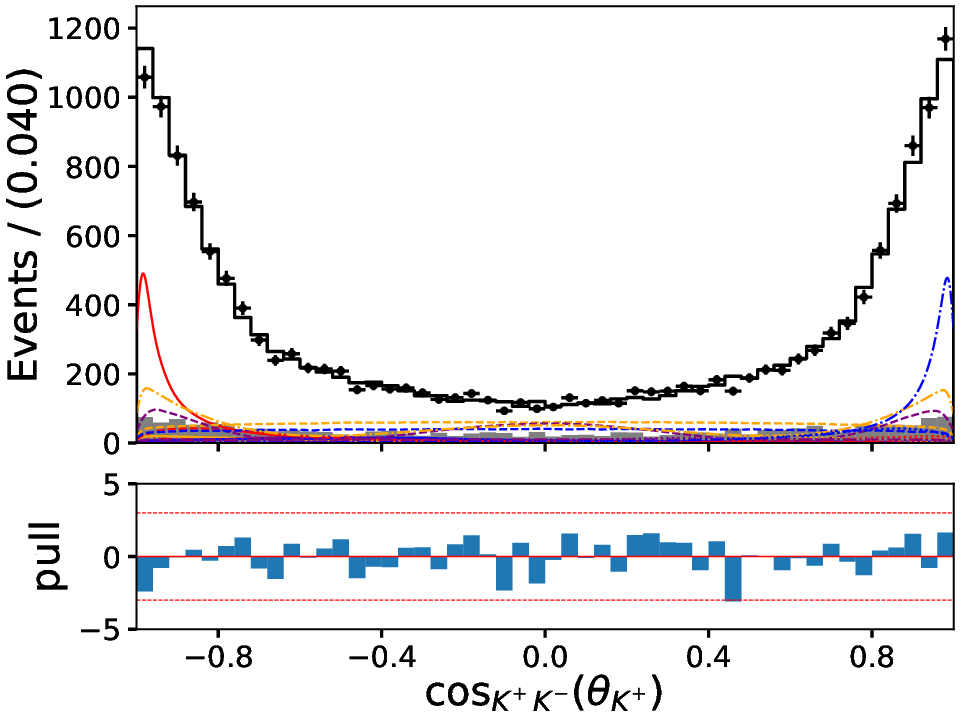}
    \put(-40,95){ \large~(e)}
    \label{fit_com_R_KpKm_Kp_cos(beta)}
    }
    \hspace{-0.1in}
    \subfigure{
    \includegraphics[width=2.in]{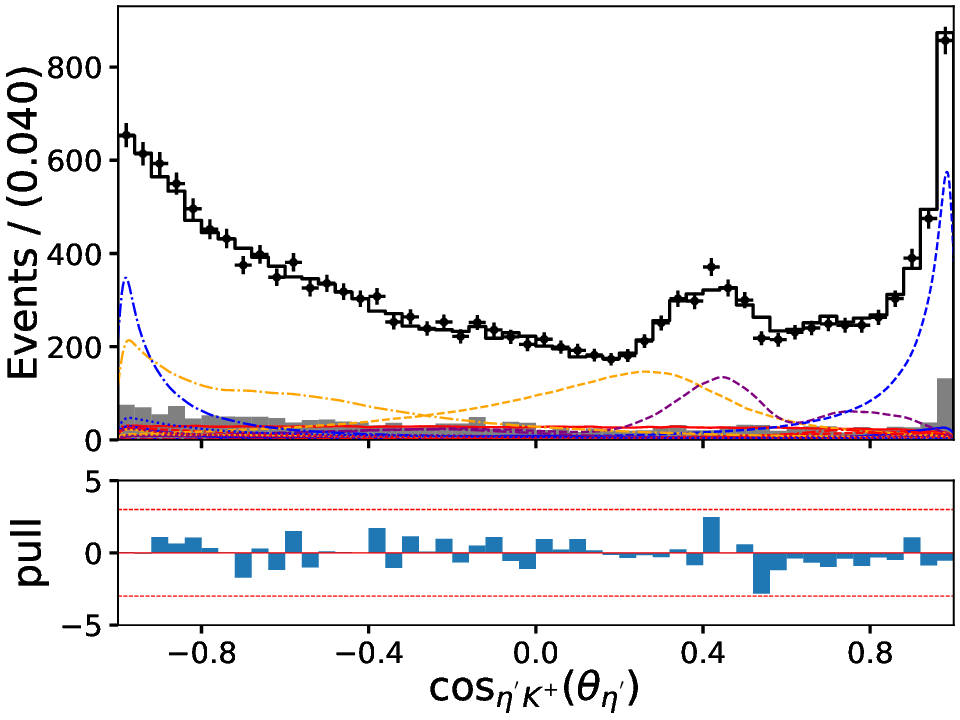}
    \put(-40,95){ \large~(f)}
    \label{fit_com_R_Kpetap_etap_cos(beta)}
    }
    \hspace{-0.1in}
\subfigure{
\raisebox{0.30in}{%
\includegraphics[width=0.7in]{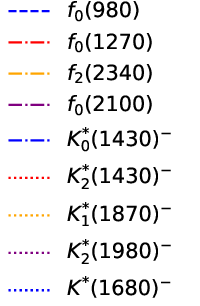}
}
}
    \caption{The  distributions of (a) $K^+K^-$,  (b) $K^+\eta^{\prime}$ and  (c) $K^-\eta^{\prime}$ invariant  masses and the angular distributions of (d) $\cos_{\chi_{c1}}(\theta_{K^+\eta^{\prime}})$, (e) $\cos_{K^+K^-}(\theta_{K^+})$ and (f) $\cos_{\eta^{\prime}K^+}(\theta_{K^+})$ for data (black dots) and the PWA fit projections (black lines). The shaded histograms are for backgrounds. The other lines represent each component. }
    \label{fit_mass_mode1}
\end{figure*}

The resulting baseline solution therefore contains a significant resonance in the $K\eta^\prime$ system with $J^{P}=1^-$, denoted as the $K_1^*(1870)$ meson, with a mass of $(1869.0 \pm 14.6)$~MeV/$c^2$ and a width of $(167.3 \pm 34.0)$~MeV. In addition, the final solution includes six resonances in the $K^+K^-$ system, namely $f_0(980)$, $f_0(1710)$, $f_0(2100)$, $f_2(1270)$, $f_2^\prime(1525)$, and $f_2(2340)$, as well as four more resonances in the $K\eta^\prime$ system, namely $K_0^*(1430)$, $K_2^*(1430)$, $K^*(1680)$, and $K_2^*(1980)$. Figure~\ref{fit_mass_mode1} shows the invariant-mass distributions of $M(K^+K^-)$, $M(K^+\eta^\prime)$, and $M(K^-\eta^\prime)$ and some angular distributions for the data, together with the projections of the PWA fit obtained from the combined analysis of the two $\eta^\prime$ decay modes. The PWA results  are summarized in Table~\ref{basic_solution}, including the masses and widths of the resonances, statistical significances, and the product branching fractions of each component. Here, the product branching fraction is defined as $\mathcal{B}(\chi_{c1}\to X^{\pm}K^{\mp}) \times \mathcal{B}(X^{\pm}\to K^{\pm}\eta^\prime)$ or $\mathcal{B}(\chi_{c1}\to Y\eta^\prime) \times \mathcal{B}(Y\to K^+K^-)$.

The measured parameters of $f_2^\prime(1525)$, $f_0(1710)$, $K_0^*(1430)$, and $K_2^*(1980)$ are consistent with the corresponding world-average values reported by the PDG~\cite{pdg}, with improved precision.  However, the coupling constants of the $K_0^*(1430)$ resonance obtained in the Flatt\'e parameterization differ from those reported in Ref.~\cite{flatte_K1430}. 
\begin{table}
\centering
\caption{The masses (M), widths (\textit{$\Gamma$}), statistical significance (Sig.) and product branching fractions ($\mathcal{B}$) for each component in the baseline solution. The first uncertainties are statistical, and the second systematic.}
\renewcommand{\arraystretch}{1.2}
\setlength{\tabcolsep}{1.0pt}
\footnotesize
\begin{tabular}{cccr}
\hline
\hline Particle  & M, \textit{$\Gamma$} (MeV/$c^{2}$, MeV) & Sig.  & $\mathcal{B}$ ($\times 10^{-5}$) \\ \hline
  $f_0(980)$    & \begin{tabular}{c} 965.0 (fixed)~\cite{f0980flatte} \\ $\mathrm{g_1}=0.165$ (fixed) \\$\mathrm{g_2} =0.695$ (fixed)\end{tabular}  & $17.1 \sigma$ & $8.7 \pm 0.9 \pm 1.3$\\ \hline
  $f_2(1270)$  & \begin{tabular}{l}1275.4 (fixed) \\185.8 (fixed) 
  \end{tabular} & $9.5 \sigma$ & $1.4 \pm 0.3 \pm 0.2$\\    \hline
  $f^{\prime}_{2}(1525)$  & \begin{tabular}{c}
       $1519.7\pm1.7\pm0.5$ \\$68.9\pm3.1\pm3.0$  
  \end{tabular}   & $32.7 \sigma$  & $10.9 \pm 0.5 \pm 0.8$\\ \hline
  $f_0(1710)$  &\begin{tabular}{c}
      $1742.0\pm4.2\pm0.9$  \\$154.8\pm7.4\pm 8.8$ 
  \end{tabular}  & $26.9 \sigma$ & $14.4 \pm 1.2 \pm 2.1$\\ \hline
  $f_0(2100)$  & \begin{tabular}{c}
       2095.0 (fixed)  \\ 287.0 (fixed)
  \end{tabular}   & $7.0 \sigma$ & $2.6 \pm 1.1 \pm 1.0$\\ \hline
  $f_2(2340)$  & \begin{tabular}{c}
      2346.0 (fixed)  \\331.0 (fixed)
  \end{tabular} & $11.0 \sigma$ & $13.1 \pm 1.2 \pm 1.5$\\ \hline
  $K_0^*(1430)^{\pm}$  & \begin{tabular}{c}$1472.2\pm4.6 \pm 7.3$ \\ $\mathrm{g}_1=0.079\pm0.017 \pm 0.011$ \\$\mathrm{g}_2=0.792\pm0.133\pm0.211$\end{tabular}  & $17.0 \sigma$ & $7.3 \pm 0.5 \pm 1.1$ \\ \hline
  $K_2^*(1430)^{\pm}$  & \begin{tabular}{c}
       1427.3 (fixed)  \\ 100.0 (fixed)
  \end{tabular} &  $9.6 \sigma$ & $1.8 \pm 0.3 \pm 0.8$ \\ \hline
  $K^*(1680)^{\pm}$    & \begin{tabular}{c}
       1718.0 (fixed)  \\320.0 (fixed)
  \end{tabular} & $11.3 \sigma$ & $2.8 \pm 0.5 \pm 0.5$ \\ \hline
  $K_2^*(1980)^{\pm}$  & \begin{tabular}{c}
       $2023.2\pm17.5\pm 22.7$  \\$295.9\pm28.8\pm 22.3$
  \end{tabular} & $14.6 \sigma$ & $2.0 \pm 0.2 \pm0.2$ \\ \hline
  $K_1^*(1870)^{\pm}$  & \begin{tabular}{c}
      $1869.0 \pm 14.6 \pm 18.6$ \\$167.3 \pm 34.0\pm 24.2$ 
  \end{tabular} &$11.0 \sigma$ &  $1.5 \pm 0.3 \pm 0.3$
 \\
\hline
\hline
\end{tabular}
\label{basic_solution}
\end{table}

\begin{table}
\centering
\renewcommand{\arraystretch}{1.2}
\caption{Relative systematic uncertainties of the branching fraction of $\chi_{c1}\to K^+K^-\eta^{\prime}$ (in \%).}
\begin{tabular}{lcc}
\hline
\hline Source & $\eta^{\prime} \rightarrow \pi^{+} \pi^{-} \gamma$ & $\eta^{\prime} \rightarrow \pi^{+} \pi^{-} \eta$ \\\hline
Tracking & 4.0 & 4.0 \\
PID & 4.0 & 2.0 \\
Photon detection & 2.0 & 3.0 \\
Kinematic fit & 1.2 & 0.3 \\
Mass window & 0.2 & 0.0 \\
Background & 0.3 & 1.0 \\
Fit procedure & 2.0 & 1.1 \\
MC model & 0.4 & 0.2 \\
Number of $\psi(2S)$ & 0.5 & 0.5 \\
Intermediate process & 3.1 & 3.0 \\
Total & 7.2 & 6.4 \\
\hline
\hline
\end{tabular}
\label{tab:sys_branch}
\end{table}

 Systematic uncertainties associated with the PWA affect both the branching fraction measurements and the determination of the resonant parameters.
The main sources include  possible  additional resonances, the resonance parameters, background estimation, Blatt-Weisskopf barrier factors~\cite{barrier_factor}, and the branching fraction of the $\chi_{c1}\to K^+K^-\eta^{\prime}$ decay. 
Uncertainties arising from possible additional resonances are evaluated by including, one at a time, the resonances with statistical significance greater than $3\sigma$ that are not part of the baseline solution.
The resulting changes in the branching fractions and resonance parameters are assigned as systematic uncertainties.
In the baseline PWA fit, the resonance parameters of the $f_0(980)$, $f_2(1270)$, $f_0(2100)$, $f_2(2340)$, $K_2^*(1430)$ and $K^*(1680)$ are fixed to their world average values~\cite{pdg}. The alternative fits are then performed in which these parameters are varied within one standard deviation, and the resulting changes are taken as systematic uncertainties.
The uncertainty due to background estimation is studied by performing alternative fits in which the background yields are varied by $\pm1\sigma$ according to their statistical uncertainties. The corresponding variations in the measured quantities are taken as systematic uncertainties.
In the PWA fit, the radius parameter in the Blatt--Weisskopf barrier factor is set to $0.6$~fm. Its associated uncertainty is estimated by varying radius parameter to $0.5$~fm and $0.7$~fm. The resulting differences in the measured masses, widths, and branching fractions are assigned as systematic uncertainties. The systematic uncertainties for each resonance parameter are summarized in
Table~\ref{basic_solution}. 

In addition,  the uncertainties associated with the tracking and PID, photon detection, kinematic fit, mass windows, fit procedure, the number of $\psi(2S)$ events~\cite{ref::psip_event}, and the branching fractions of $\psi(2S)\to\gamma\chi_{c1}$ and $\eta^{\prime}\to\pi^+\pi^-\gamma(\eta\to\gamma\gamma)$ from the PDG~\cite{pdg}, are also taken into account in the measurement of branching fractions. 
The uncertainties on the tracking and PID for each charged track and the photon selection are investigated using the control samples of $J/\psi\to K^{*}\bar{K}$~\cite{sys_tracking_K} or $ J/\psi\to \rho\pi$~\cite{sys_tracking_pi}, 
$J/\psi\to K^+K^-K^0$ or $ J/\psi\to \pi^+\pi^-\pi^0$~\cite{sys_pid}, and $J/\psi\to \rho \pi^0$~\cite{ref:photon}, respectively. 
The differences between data and MC simulation are 2.0\% per charged track including the tracking and PID and 1.0\% per photon. 
The uncertainties associated with the  kinematic fits are estimated by correcting the track helix parameters. The differences 
with and without the correction are taken as the systematic uncertainties.
The systematic uncertainties of $\eta^{\prime}$ mass windows are estimated from the data--MC difference in the $\eta^{\prime}$ mass resolution, obtained by fitting the data $\eta^{\prime}$ mass spectrum with the MC signal shape convolved with a Gaussian and comparing the mass window efficiencies.
Alternative fits are performed with different Chebyshev functions, $\eta^{\prime}$ sideband regions, and signal shapes. The largest differences for each item are added in quadrature and taken as a systematic uncertainty.
An alternative amplitude model is used to generate the signal MC sample, and the efficiency difference from the nominal one is also taken as a systematic uncertainty.
Assuming all of these sources are independent, the total systematic uncertainties are 7.5\% and 6.5\% for $\pi^+\pi^-\gamma$ and $\pi^+\pi^-\eta$, respectively, which are summarized in Table~\ref{tab:sys_branch}.

In summary, based on a sample of $(2712.4\pm14.3)\times10^6$ $\psi(2S)$ events collected with the BESIII detector, a combined PWA of $\psi(2S)\to \gamma\chi_{c1}, \chi_{c1}\to K^+K^-\eta^{\prime}$ with $\eta^{\prime}\to \pi^+\pi^-\gamma$ and $\eta^{\prime}\to \pi^+\pi^-\eta$ is performed. A resonant state with $J^P=1^-$ in the $K\eta^{\prime}$ system, denoted as the $K_1^*(1870)$ meson, is observed for the first time with a statistical significance exceeding 10$\sigma$. Its mass and width are measured to be $M=(1869.0 \pm 14.6\pm 18.6)$ MeV/$c^{2}$ and $\Gamma=(167.3 \pm 34.0 \pm 24.2)$ MeV. The branching fraction of $\chi_{c1}\to K^{\mp}K_1^*(1870)^{\pm}\to K^{+}K^{-}\eta^{\prime}$ is determined to be $(1.48 \pm 0.30 \pm 0.33)\times 10^{-5}$. 
Theoretically, Barnes et al.~\cite{K1950} used the $^{3}P_0$ model to predict a $3^{3}S_1$ kaon state with a mass around 1950 MeV/$c^{2}$, referred to as  $K^{*}(1950)$. Its dominant decay mode is expected to be $\rho K^*$, while the decays to  $\eta K$ and  $\eta^{\prime} K$ are predicted to be suppressed. 
In addition,  Feng et al.~\cite{K1950_2} estimated the mass of $K(3^{3}S_1)$ from Regge trajectories with two different slopes, 1.95 or 2.23 GeV/$c^{2}$, with the corresponding total widths of 176 or 760 MeV, respectively. Since an extremely large width of about 700 MeV is unlikely for a resonant structure observed in this analysis, the more plausible expectation for $K(3^{3}S_1)$ would be a mass near 2 GeV/$c^{2}$ and a width of about $200$ MeV. However,
the predicted masses in both theoretical scenarios are not consistent with our measured result, which makes a hybrid interpretation a plausible possibility. Nevertheless, further theoretical and experimental studies are still needed to clarify the nature of the resonant structure observed in the $K\eta^\prime$ invariant mass spectrum.

In addition to the newly observed $K_1^*(1870)$ meson, the intermediate processes $\chi_{c1}\to f_{0}(980)\eta^\prime$, $f_{0}(1710)\eta^\prime$, $f_{0}(2100)\eta^\prime$, $f_{2}(1270)\eta^\prime$, $f_2^\prime(1525)\eta^\prime$, $f_{2}(2340)\eta^\prime$, $K_0^*(1430)^{\pm}K^{\mp}$, $K_2^*(1430)^{\pm}K^{\mp}$, $K^*(1680)^{\pm}K^{\mp}$, and $K_2^*(1980)^{\pm}K^{\mp}$ are found to be necessary to describe the data, each with a statistical significance greater than $5\sigma$. The corresponding branching fractions are determined and summarized in Table~\ref{basic_solution}.
In Ref.~\cite{chic1}, the branching fraction of $\chi_{c1}\to f_{0}(1710)\eta^\prime$ is calculated under different glueball--$q\bar q$ mixing schemes. Using the PDG value for the branching fraction of $f_{0}(1710)\to K^+K^-$~\cite{pdg}, we determine the branching fraction of $\chi_{c1}\to f_{0}(1710)\eta^\prime$ to be $(4.01\pm1.50)\times10^{-4}$. This result is consistent within $2\sigma$ with Scheme~I of the glueball--$q\bar q$ mixing scenarios discussed in Ref.~\cite{chic1}. In that scheme, the $f_{0}(1710)$ is dominated by the $s\bar{s}$ component, although it also contains a sizable glueball contribution. However, because of the relatively large uncertainty in $\mathcal{B}(f_{0}(1710)\to K^+K^-)$ quoted by the PDG~\cite{pdg}, our result is also consistent with Scheme~II and Scheme~III within $3\sigma$. Finally, the combined branching fraction of $\chi_{c1}\to K^+K^-\eta^\prime$ is measured to be $(7.90 \pm 0.48)\times10^{-4}$, which is consistent with, and supersedes, the previous BESIII measurement~\cite{previouswork}.


The BESIII Collaboration thanks the staff of BEPCII (https://cstr.cn/31109.02.BEPC) and the IHEP computing center for their strong support. This work is supported in part by National Key R\&D Program of China under Contracts Nos. 2025YFA1613900, 2023YFA1606000, 2023YFA1606704; National Natural Science Foundation of China (NSFC) under Contracts Nos. 11635010, 11935015, 11935016, 11935018, 12025502, 12035009, 12035013, 12061131003, 12192260, 12192261, 12192262, 12192263, 12192264, 12192265, 12221005, 12225509, 12235017, 12342502, 12361141819, 12535005; the Chinese Academy of Sciences (CAS) Large-Scale Scientific Facility Program; the Strategic Priority Research Program of Chinese Academy of Sciences under Contract No. XDA0480600; CAS under Contract No. YSBR-101; 100 Talents Program of CAS; The Institute of Nuclear and Particle Physics (INPAC) and Shanghai Key Laboratory for Particle Physics and Cosmology; Agencia Nacional de Investigaci\'on y Desarrollo de Chile (ANID), Chile under Contract No. ANID CCTVal CIA250027; Istituto Nazionale di Fisica Nucleare, Italy; Knut and Alice Wallenberg Foundation under Contracts Nos. 2021.0174, 2021.0299, 2023.0315; Ministry of Development of Turkey under Contract No. DPT2006K-120470; National Research Foundation of Korea under Contract No. RS-2026-25486791; National Science and Technology fund of Mongolia; Polish National Science Centre under Contract No. 2024/53/B/ST2/00975; STFC (United Kingdom); Swedish Research Council under Contract No. 2019.04595; U. S. Department of Energy under Contract No. DE-FG02-05ER41374


\bibliographystyle{apsrev4-1}
\bibliography{references}

@article{Meyer:2015gp,
	title = {Hybrid mesons},
	journal = {Prog. Part. Nucl. Phys.},
	volume = {82},
	pages = {21-58},
	year = {2015},
	issn = {0146-6410},
	doi = {https://doi.org/10.1016/j.ppnp.2015.03.001},
	url = {https://www.sciencedirect.com/science/article/pii/S0146641015000162},
	author = {C.A. Meyer and E.S. Swanson},
}

@article{KLEMPT20071,
    title = {Glueballs, hybrids, multiquarks: Experimental facts versus QCD inspired concepts},
    journal = {Phys. Rept.},
    volume = {454},
    number = {1},
    pages = {1-202},
    year = {2007},
    issn = {0370-1573},
    doi = {https://doi.org/10.1016/j.physrep.2007.07.006},
    url = {https://www.sciencedirect.com/science/article/pii/S0370157307003560},
    author = {Eberhard Klempt and Alexander Zaitsev},
}

@article{BESIIIdetector,
    author = "Ablikim, M. and others",
    collaboration = "BESIII",
    title = "{Design and construction of the BESIII detector}",
    doi = "10.1016/j.nima.2009.12.050",
    journal = "Nucl. Instrum. Meth. A",
    volume = "614",
    pages = "345--399",
    year = "2010"
}

@article{geant4,
    author = "Agostinelli, S. and others",
    collaboration = "GEANT4 Collaboration",
    title = "{GEANT4--a simulation toolkit}",
    reportNumber = "SLAC-PUB-9350, FERMILAB-PUB-03-339, CERN-IT-2002-003",
    doi = "10.1016/S0168-9002(03)01368-8",
    journal = "Nucl. Instrum. Meth. A",
    volume = "506",
    pages = "250--303",
    year = "2003"
}

@article{pdg,
author = { Takahashi, F. and others},
collaboration = {Particle Data Group},
title = {Review of Particle Physics*},
journal = {Int. J. Mod. Phys. A},
volume = {41},
number = {22},
pages = {2630011},
year = {2026},
doi = {10.1142/S0217751X26300115},
}

@article{ref::psip_event,
   author = {M. Ablikim and others},
   collaboration = {BESIII Collaboration},
   title = {Determination of the number of ψ(3686) events taken at BESIII*},
   journal = {Chin. Phys. C},
   year = {2024},
   month = {sep},
   volume = {48},
   number = {9},
   pages = {093001},
   doi = {10.1088/1674-1137/ad595b},
   url = {https://dx.doi.org/10.1088/1674-1137/ad595b},
}

@article{sys_tracking_K,
  title = {Observation of the decays ${\ensuremath{\chi}}_{cJ}\ensuremath{\rightarrow}{\mathrm{\ensuremath{\Sigma}}}^{0}\overline{p}{K}^{+}+\mathrm{c}.\mathrm{c}.$ ($J=0,1,2$)},
  author = {Ablikim, M. and others},
  collaboration = {BESIII Collaboration},
  journal = {Phys. Rev. D},
  volume = {102},
  issue = {9},
  pages = {092006},
  numpages = {10},
  year = {2020},
  month = {Nov},
  publisher = {American Physical Society},
  doi = {10.1103/PhysRevD.102.092006},
  url = {https://link.aps.org/doi/10.1103/PhysRevD.102.092006}
}

@article{sys_tracking_pi,
  title = {Observation of $\ensuremath{\eta}\ensuremath{'}\ensuremath{\rightarrow}{\ensuremath{\pi}}^{+}{\ensuremath{\pi}}^{\ensuremath{-}}{\ensuremath{\pi}}^{+}{\ensuremath{\pi}}^{\ensuremath{-}}$ and $\ensuremath{\eta}\ensuremath{'}\ensuremath{\rightarrow}{\ensuremath{\pi}}^{+}{\ensuremath{\pi}}^{\ensuremath{-}}{\ensuremath{\pi}}^{0}{\ensuremath{\pi}}^{0}$},
  author = {Ablikim, M. and others},
  collaboration = {BESIII Collaboration},
  journal = {Phys. Rev. Lett.},
  volume = {112},
  issue = {25},
  pages = {251801},
  numpages = {6},
  year = {2014},
  month = {Jun},
  publisher = {American Physical Society},
  doi = {10.1103/PhysRevLett.112.251801},
  url = {https://link.aps.org/doi/10.1103/PhysRevLett.112.251801}
}

@article{sys_pid,
  title = {Study of ${\ensuremath{\chi}}_{cJ}$ radiative decays into a vector meson},
  author = {Ablikim, M. and others},
  collaboration = {BESIII Collaboration},
  journal = {Phys. Rev. D},
  volume = {83},
  issue = {11},
  pages = {112005},
  numpages = {9},
  year = {2011},
  month = {Jun},
  publisher = {American Physical Society},
  doi = {10.1103/PhysRevD.83.112005},
  url = {https://link.aps.org/doi/10.1103/PhysRevD.83.112005}
}

@article{chic1,
author = {Wang, QIAN and Li, GANG and Zhao, QIANG},
title = {PROBING SCALAR MESON STRUCTURES IN χc1 DECAYS INTO PSEUDOSCALAR AND SCALAR},
journal = {Int. J. Mod. Phys. A},
volume = {27},
number = {23},
pages = {1250135},
year = {2012},
doi = {10.1142/S0217751X12501357},
URL = { 
        https://doi.org/10.1142/S0217751X12501357
},
}

@article{lightest_hybrid,
  title = {Status of exotic-quantum-number mesons},
  author = {Meyer, C. A. and Van Haarlem, Y.},
  journal = {Phys. Rev. C},
  volume = {82},
  issue = {2},
  pages = {025208},
  numpages = {20},
  year = {2010},
  month = {Aug},
  publisher = {American Physical Society},
  doi = {10.1103/PhysRevC.82.025208},
  url = {https://link.aps.org/doi/10.1103/PhysRevC.82.025208}
}

@article{pi1600,
  title = {Amplitude analyses of the decays ${\ensuremath{\chi}}_{c1}\ensuremath{\rightarrow}\ensuremath{\eta}{\ensuremath{\pi}}^{+}{\ensuremath{\pi}}^{\ensuremath{-}}$ and ${\ensuremath{\chi}}_{c1}\ensuremath{\rightarrow}{\ensuremath{\eta}}^{\ensuremath{'}}{\ensuremath{\pi}}^{+}{\ensuremath{\pi}}^{\ensuremath{-}}$},
  author = {Adams, G. S. and others},
  collaboration = {CLEO Collaboration},
  journal = {Phys. Rev. D},
  volume = {84},
  issue = {11},
  pages = {112009},
  numpages = {17},
  year = {2011},
  month = {Dec},
  publisher = {American Physical Society},
  doi = {10.1103/PhysRevD.84.112009},
  url = {https://link.aps.org/doi/10.1103/PhysRevD.84.112009}
}

@article{eta1855,
  title = {Observation of an Isoscalar Resonance with Exotic ${J}^{PC}={1}^{\ensuremath{-}+}$ Quantum Numbers in $J/\ensuremath{\psi}\ensuremath{\rightarrow}\ensuremath{\gamma}\ensuremath{\eta}{\ensuremath{\eta}}^{\ensuremath{'}}$},
  author = {Ablikim, M. and others},
  collaboration = {BESIII Collaboration},
  journal = {Phys. Rev. Lett.},
  volume = {129},
  issue = {19},
  pages = {192002},
  numpages = {9},
  year = {2022},
  month = {Oct},
  publisher = {American Physical Society},
  doi = {10.1103/PhysRevLett.129.192002},
  url = {https://link.aps.org/doi/10.1103/PhysRevLett.129.192002}
}

@article{previouswork,
  title = {Measurement of ${\ensuremath{\chi}}_{cJ}$ decaying into ${\ensuremath{\eta}}^{\ensuremath{'}}{K}^{+}{K}^{\ensuremath{-}}$},
  author = {Ablikim, M. and others},
  collaboration = {BESIII Collaboration},
  journal = {Phys. Rev. D},
  volume = {89},
  issue = {7},
  pages = {074030},
  numpages = {18},
  year = {2014},
  month = {Apr},
  publisher = {American Physical Society},
  doi = {10.1103/PhysRevD.89.074030},
  url = {https://link.aps.org/doi/10.1103/PhysRevD.89.074030}
}

@article{barrier_factor,
  title = {Centrifugal-Barrier Effects in Resonance Partial Decay Widths, Shapes, and Production Amplitudes},
  author = {von Hippel, Frank and Quigg, C.},
  journal = {Phys. Rev. D},
  volume = {5},
  issue = {3},
  pages = {624--638},
  numpages = {0},
  year = {1972},
  month = {Feb},
  publisher = {American Physical Society},
  doi = {10.1103/PhysRevD.5.624},
  url = {https://link.aps.org/doi/10.1103/PhysRevD.5.624}
}

@article{ref:photon,
    author = "Ablikim, M. and others",
    collaboration = "BESIII Collaboration",
    title = "{Branching fraction measurements of $\chi_{c0}$ and $\chi_{c2}$ to $\pi^0\pi^0$ and $\eta\eta$}",
    primaryClass = "hep-ex",
    doi = "10.1103/PhysRevD.81.052005",
    journal = "Phys. Rev. D",
    volume = "81",
    pages = "052005",
    year = "2010"
}

@article{TF-PWA,
  title = {open source framework TF-PWA package},
  author = {Yi Jiang and others},
  journal = {https://github.com/jiangyi15/tf-pwa},
  volume = {},
  issue = {},
  pages = {},
  numpages = {},
  year = {2020},
  month = {},
  publisher = {},
  url = {https://github.com/jiangyi15/tf-pwa}
}

@article{flatte,
title = {Coupled-channel analysis of the πη and KK̄ systems near KK̄ threshold},
journal = {Phys. Lett. B},
volume = {63},
number = {2},
pages = {224-227},
year = {1976},
issn = {0370-2693},
doi = {https://doi.org/10.1016/0370-2693(76)90654-7},
url = {https://www.sciencedirect.com/science/article/pii/0370269376906547},
author = {S.M. Flatté},
}

@article{f0980flatte,
title = {Resonances in J/ψ→ϕπ+π− and ϕK+K−},
author = {M. Ablikim and others},
collaboration = {BESIII Collaboration},
journal = {Phys. Lett. B},
volume = {607},
number = {3},
pages = {243-253},
year = {2005},
issn = {0370-2693},
doi = {https://doi.org/10.1016/j.physletb.2004.12.041},
url = {https://www.sciencedirect.com/science/article/pii/S0370269304017265},
}

@article{K1950,
  title = {Strong decays of strange quarkonia},
  author = {Barnes, T. and Black, N. and Page, P. R.},
  journal = {Phys. Rev. D},
  volume = {68},
  issue = {5},
  pages = {054014},
  numpages = {43},
  year = {2003},
  month = {Sep},
  publisher = {American Physical Society},
  doi = {10.1103/PhysRevD.68.054014},
  url = {https://link.aps.org/doi/10.1103/PhysRevD.68.054014}
}

@article{K1950_2,
  title = {Possible assignment of excited light ${^{3}S}_{1}$ vector mesons},
  author = {Feng, Jie Cheng and Kang, Xian Wei and L\"u, Qi Fang and Zhang, Feng Shou},
  journal = {Phys. Rev. D},
  volume = {104},
  issue = {5},
  pages = {054027},
  numpages = {11},
  year = {2021},
  month = {Sep},
  publisher = {American Physical Society},
  doi = {10.1103/PhysRevD.104.054027},
  url = {https://link.aps.org/doi/10.1103/PhysRevD.104.054027}
}

@article{hybrid,
title = {Hybrid mesons from quenched QCD},
journal = {Phys. Lett. B},
volume = {401},
number = {3},
pages = {308-312},
year = {1997},
issn = {0370-2693},
url = {https://www.sciencedirect.com/science/article/pii/S0370269397003845},
author = {P. Lacock and C. Michael and P. Boyle and P. Rowland}
}

@article{pi1400,
  author = {Thompson, D. R. and others},
  collaboration = {E852 Collaboration},
  journal = {Phys. Rev. Lett.},
  volume = {79},
  issue = {9},
  pages = {1630--1633},
  numpages = {0},
  year = {1997},
  month = {Sep},
  publisher = {American Physical Society},
  doi = {10.1103/PhysRevLett.79.1630},
  url = {https://link.aps.org/doi/10.1103/PhysRevLett.79.1630}
}

@article{pi2015,
title = {Exotic meson production in the f1(1285)π− system observed in the reaction π−p→ηπ+π−π−p at 18 GeV/c},
journal = {Phys. Lett. B},
volume = {595},
number = {1},
pages = {109-117},
year = {2004},
issn = {0370-2693},
doi = {https://doi.org/10.1016/j.physletb.2004.05.032},
url = {https://www.sciencedirect.com/science/article/pii/S0370269304007798},
author = {J. Kuhn and others}
}

@Article{nonet_1,
author = {Lin Qiu and Qiang Zhao},
title = {Towards the establishment of the light <i>J</i><sup><i>P</i> (<i>C</i> )</sup>=1<sup>–(+)</sup> hybrid nonet},
doi = {10.1088/1674-1137/ac567e},
journal = {Chin. Phys. C},
volume = {46},
number = {5},
pages = {051001},
year = {2022},
url = {https://hepnp.ihep.ac.cn/en/article/doi/10.1088/1674-1137/ac567e},
}

@article{nonet_2,
  title = {Constructing the ${J}^{P(C)}={1}^{\ensuremath{-}(+)}$ light flavor hybrid nonet with the newly observed ${\ensuremath{\eta}}_{1}(1855)$},
  author = {Chen, Bing and Luo, Si Qiang and Liu, Xiang},
  journal = {Phys. Rev. D},
  volume = {108},
  issue = {5},
  pages = {054034},
  numpages = {11},
  year = {2023},
  month = {Sep},
  publisher = {American Physical Society},
  doi = {10.1103/PhysRevD.108.054034},
  url = {https://link.aps.org/doi/10.1103/PhysRevD.108.054034}
}

@article{flatte_K1430,
  title = {Dalitz plot analysis of the ${D}^{+}\ensuremath{\rightarrow}{K}^{\ensuremath{-}}{\ensuremath{\pi}}^{+}{\ensuremath{\pi}}^{+}$ decay},
  author = {Bonvicini, G. and others},
  collaboration = {CLEO Collaboration},
  journal = {Phys. Rev. D},
  volume = {78},
  issue = {5},
  pages = {052001},
  numpages = {20},
  year = {2008},
  month = {Sep},
  publisher = {American Physical Society},
  doi = {10.1103/PhysRevD.78.052001},
  url = {https://link.aps.org/doi/10.1103/PhysRevD.78.052001}
}

@article{BESIII:2026ebb,
    author = "Ablikim, Medina and others",
    collaboration = "BESIII Collaboration",
    title = "{Observation of the Exotic State $π_{1}(1600)$ in $ψ(2S)\rightarrowγχ_{c1},χ_{c1}\rightarrowπ^{+}π^{-}η'$}",
    eprint = "2604.12524",
    archivePrefix = "arXiv",
    month = "4",
    year = "2026"
}

\end{document}